\documentclass[11pt,a4paper]{article}

\usepackage[T1]{fontenc}
\usepackage[utf8]{inputenc}
\usepackage{lmodern}
\usepackage[a4paper,margin=2.4cm]{geometry}
\usepackage{graphicx}
\usepackage{textcomp}
\usepackage{eurosym}
\usepackage{xcolor}
\usepackage{titlesec}
\usepackage{abstract}
\usepackage{microtype}
\usepackage{enumitem}
\usepackage[round,authoryear]{natbib}
\usepackage[colorlinks=true,linkcolor=black,citecolor=blue!50!black,urlcolor=blue!50!black]{hyperref}
\usepackage{caption}

\graphicspath{{figures/}}

\titleformat{\section}{\normalfont\large\bfseries}{\thesection.}{0.6em}{}
\titleformat{\subsection}{\normalfont\bfseries}{\thesubsection.}{0.6em}{}

\begin{document}

\noindent{\small\sffamily\bfseries\MakeUppercase{Translational Article}}\par
\vspace{0.5em}
\hrule
\vspace{1em}

\begin{center}
{\LARGE\bfseries From Licensing to Open Access:\\Designing a Sustainable Transition in Operational Weather Data\par}
\vspace{1.2em}
{\large Emma Pidduck\textsuperscript{1,*}, Umberto Modigliani\textsuperscript{1}, Victoria Bennett\textsuperscript{1},\\Fabio Venuti\textsuperscript{2}, Florian Pappenberger\textsuperscript{2}, Florence Rabier\textsuperscript{3}\par}
\vspace{1em}
{\small
\textsuperscript{1}Forecasts \& Services Department, European Centre for Medium-Range Weather Forecasts (ECMWF), Bonn, Germany\\
\textsuperscript{2}Office of the Director General, European Centre for Medium-Range Weather Forecasts (ECMWF), Reading, RG2 9AX, Berkshire, United Kingdom\\
\textsuperscript{3}Non-Executive Director, Met Office, United Kingdom\\[0.5em]
\textsuperscript{*}Corresponding author: \href{mailto:emma.pidduck@ecmwf.int}{emma.pidduck@ecmwf.int}
}
\end{center}

\vspace{1em}


\begin{abstract}
\noindent This translational article documents the European Centre for Medium-Range Weather Forecasts (ECMWF) transition from a restricted data licensing model to open access under CC BY 4.0, completed in October 2025. The policy context included EU open data requirements and alignment with international data exchange frameworks. The transition was implemented through a tiered service model that kept core forecast data open while offering operationally supported delivery as a cost-recovered service. Between 2020 and 2025, ECMWF executed an iterative planning cycle: setting an annual target for revenue reduction, specifying additions to the open tier under that target, provisioning infrastructure, and assessing outcomes to update assumptions. Drawing on internal administrative records (2014--2025), we describe design choices, operational constraints, and early outcomes. In the six months following the end of the transition, more than 93\% of previously paying organisations retained a Service Agreement, while open endpoint download volumes increased substantially. We discuss trade-offs in defining the open tier (resolution, parameters, schedule), the reduction of compliance overheads formerly associated with redistribution restrictions, and the scalability implications of global distribution. We note an emerging sustainability question as AI-based forecast products become freely available. The early evidence is consistent with the view that a tiered service model can be designed to reconcile open-access obligations with operational sustainability, subject to monitoring over longer contract renewal cycles (typically annual).
\end{abstract}

\bigskip

\noindent\textbf{Policy Significance Statement.} Public data institutions frequently face a practical dilemma: how to meet open-access expectations without undermining the operational services that produce and deliver the data. This article describes ECMWF's transition (2020--2025) from a restricted licensing model to open access under CC BY 4.0, implemented through a tiered model that separates free access to core data from cost-recovered operational delivery. The case links policy obligations to design decisions under real-world infrastructure constraints and reports early outcomes over six months. It identifies design elements with wider applicability---scope definition aligned with international standards, service differentiation focused on reliability and support, and investment in self-service tooling. The evidence suggests that institutions can expand access to public data while maintaining service viability, provided transparent metrics and scalable delivery are resourced.

%%%%%%%%%%%%%%%%%%%%%%%%%%%%%%%%%%%%%%%%%%%%%%%%%%%%%%%%%%%%%%%%
\section{Introduction}\label{sec:intro}

Open data is a catalyst for scientific progress, economic innovation, and informed decision-making across both public and private sectors. The economic case is well established: studies consistently demonstrate that open data reduces duplication across public institutions, accelerates research and development, and enables private sector innovation across sectors from agricultural technology and energy trading to logistics and financial risk modelling \citep{janssen2012}. The McKinsey Global Institute estimated that open data could unlock more than \$3 trillion in additional economic value annually across seven sectors of the global economy \citep{manyika2013}, while the European Commission projected the direct economic value of public sector information reuse to rise from \euro52 billion in 2018 to \euro194 billion by 2030 \citep{ec2020}. Sector-specific evidence reinforces these estimates: the elimination of data sales charges for the Landsat satellite archive in 2008 increased daily downloads by more than 100-fold and generated an estimated annual economic benefit exceeding the foregone revenue by two orders of magnitude \citep{zhu2019}.

Meteorology and climatology provide clear examples of these benefits, where the free exchange of observations, model outputs, and climate records underpins the early warning systems, emergency response frameworks, adaptation strategies, and economic growth on which societies depend. For the first time in recorded history, the world has experienced three consecutive years with a global average temperature exceeding 1.5\textdegree C above pre-industrial levels \citep{c3s2026}. In 2025 alone, natural disasters claimed at least 42{,}000 lives and caused an estimated \$260 billion in economic losses globally, including more than 24{,}500 deaths in Europe \citep{aon2026}. These trends increase the practical importance of accessible meteorological data for science, policy, and operational decision-making. More recently, open data has become a vital input for artificial intelligence and machine learning systems, which are increasingly central to modern forecasting, climate modelling, and decision support \citep{edp2020}.

The tension between open access and institutional sustainability is not easily resolved. For institutions that depend on, or are supplemented by, licensing revenue, transitioning to, or increasing the availability of, free and open data risks destabilising the financial model that enables high-quality forecasting. The European Centre for Medium-Range Weather Forecasts (ECMWF), an independent intergovernmental organisation supported by 35 Member and Co-operating States, faced exactly this challenge. Although not an EU body, the majority of ECMWF's Member States are EU members, and between 2008 and 2025 it generated approximately \euro135 million in cumulative licensing revenue from data access agreements. This model that increasingly sat in tension with open science and open data expectations embedded in EU policy, most notably the Open Data Directive \citep{eu2019} and the European Strategy for Data \citep{ec2020}. In this paper, we examine a tiered service model, or ``freemium'' model, in which data are open while operational delivery is cost-recovered and assess whether this arrangement supports open access without undermining service provision. The goal is not to treat open data as a residual offering but to make it as useful as possible within the constraints of what can be reliably and affordably delivered to a global user base. In some national contexts, comparable transitions have been supported by additional public investment: Denmark's ``Free Data'' initiative, launched as part of the Digital Growth Strategy, funded the progressive release of Danish Meteorological Institute (DMI) data under CC BY 4.0 via the Danish GovCloud from 2020 onwards \citep{iid2020}. ECMWF's transition proceeded without a dedicated supplementary budget, relying instead on a tiered service model for financial sustainability.

When the open data policy was fully implemented, in October 2025, ECMWF released its entire operational forecast data catalogue under a Creative Commons Attribution 4.0 International (CC-BY-4.0) licence. This paper documents that transition with particular attention to the tiered service model that made it financially viable. Drawing on internal administrative records covering the period 2014 to 2025, we show that user uptake expanded and revenue was retained through enhanced service offerings rather than through data restrictions, while significantly increasing the volume of free and open data. The lessons are applicable beyond meteorology: any public data institution navigating the shift from cost-recovery to open provision will face analogous pressures \citep{zuiderwijk2014}, and the approach ECMWF adopted offers a transferable framework for doing so without sacrificing the operational capacity on which data quality depends. The framework is likely to transfer most readily to markets that share key features with operational meteorology: demand that is sensitive to timeliness and reliability, a demonstrable economic advantage from real-time access, and limited globally available substitutes.

%%%%%%%%%%%%%%%%%%%%%%%%%%%%%%%%%%%%%%%%%%%%%%%%%%%%%%%%%%%%%%%%
\section{Background: The Prior Licensing Model and Policy Context}\label{sec:background}

For much of its operational history, ECMWF generated revenue from data licensing through a framework developed in partnership with its Member and Co-operating States, whose National Meteorological and Hydrological Services (NMHSs) are ECMWF's primary operational partners and the principal recipients of its forecast data. The main objective of this licensing framework was to ensure that the costs of producing and distributing high-quality meteorological data were shared appropriately, while preserving the authoritative role of Member States as primary distributors within their territories.

At the core of this model was the ECOMET Product Unit (EPU)---a standardised unit cost applied to meteorological parameters, originally designed for pricing data distributed through ECOMET (the Economic Interest Grouping of the European NMHSs; see \url{www.ecomet.eu}), now part of EUMETNET. ECMWF adopted the EPU as the basis for its own ``Information Cost'' calculations, creating a pricing structure in which users pay based on the number of meteorological parameters, the geographical area of interest, and the resolution of the data provided. ``Information Cost'' became the `value' of data, and delivery was subject to further charges, which were established by the distributor. Annual agreements were negotiated individually, with Member and Co-operating States also holding the right to sublicense ECMWF data within and outside their territories.

It is important to contextualise the origins of this model. ECMWF did not develop data licensing independently; it did so at the explicit request of its Member and Co-operating States in the 1980s, for whom the arrangement served several legitimate purposes. The licensing framework provided a mechanism to generate returns on publicly funded infrastructure while preserving the authoritative position of Member States as primary distributors of meteorological data within their territories. As a not-for-profit Intergovernmental Organisation, ECMWF does not generate surplus for distribution; all licensing revenue was reinvested directly into ECMWF's annual operational budget, funding the computational infrastructure, scientific development, and staff capacity that underpin forecast quality and delivery.

The licensing terms were periodically revised to reflect changing technological, political and commercial conditions. A central provision of the original policy was the prohibition on redistribution by licensed users. The licence distinguished between ``Retrievable Value Added Services'' (R-VAS)---services from which the underlying ECMWF data could be retrieved or reverse-engineered without significant effort---and ``Non-retrievable Value Added Services'' (N-VAS), from which the underlying data could not be extracted. R-VAS could be distributed only to clearly identified third parties under controlled conditions; N-VAS, such as graphical products derived from ECMWF data, could be shared more broadly. In practice, the boundary between legitimate value addition and redistribution became a persistent source of tension and administrative overhead. Users tested the limits of compliance, and cases emerged in which minimal data transformations were applied solely to circumvent the no-redistribution clause. In some cases, commercial actors built substantive business models on data presented as ECMWF-derived, applying transformations of questionable materiality to satisfy the letter of the redistribution prohibition while arguably circumventing its intent. The eventual resolution of such cases required establishing a principle that remains underappreciated: genuinely transformed data is no longer the originating provider's data and cannot legitimately be marketed as such. Notably, enforcement relied less on ECMWF's formal policing than on users' competitive vigilance, which had strong commercial incentives to report suspected violations by rivals. This compliance burden nonetheless consumed significant administrative resources and generated ongoing friction with the user community.

It is also notable that the Retrievable/Non-retrievable Value Added Service distinction, a central component of ECMWF's licensing framework for several decades, has since been adopted by Google. Google's terms of use for its WeatherNext real-time experimental data \citep{google2025} replicate this architecture almost precisely: distinguishing between Retrievable and Non-Retrievable Value Added Services, applying the same logic that more freely shareable outputs are those from which the underlying data cannot be recovered, and explicitly enumerating transformations (colouring, formatting, compressing, sub-setting) that do not constitute sufficient modification to qualify as a Value Added Service. That a framework developed in the 1990s to manage commercial redistribution of NWP output has been adopted for AI forecast products in 2025 suggests it captures something structurally important about the problem of governing high-value meteorological data in competitive markets.

The financial model also evolved over time, driven both by technological change and by the strategic direction toward open data. Initially, an additional ``Handling Charge'' equivalent to 20\% of the Information Cost was applied to cover data-delivery costs, as these were manually determined. However, each Licensor (Member and Co-operating State) could also define their own Handling Charges if distributing via its own infrastructure. As technology evolved, ECMWF began introducing self-service tools in approximately 2018 that shifted responsibility for data access configuration to the user. In 2022, the Handling Charge was formally replaced by a structured ``Service Charge'' scheme. Separately, and reflecting the deliberate path toward open access, the Information Cost was placed on a downward trajectory from 2022, with the intention of reaching zero by the time of the completed transition.

\begin{figure}[htbp]
\centering
\includegraphics[width=\linewidth]{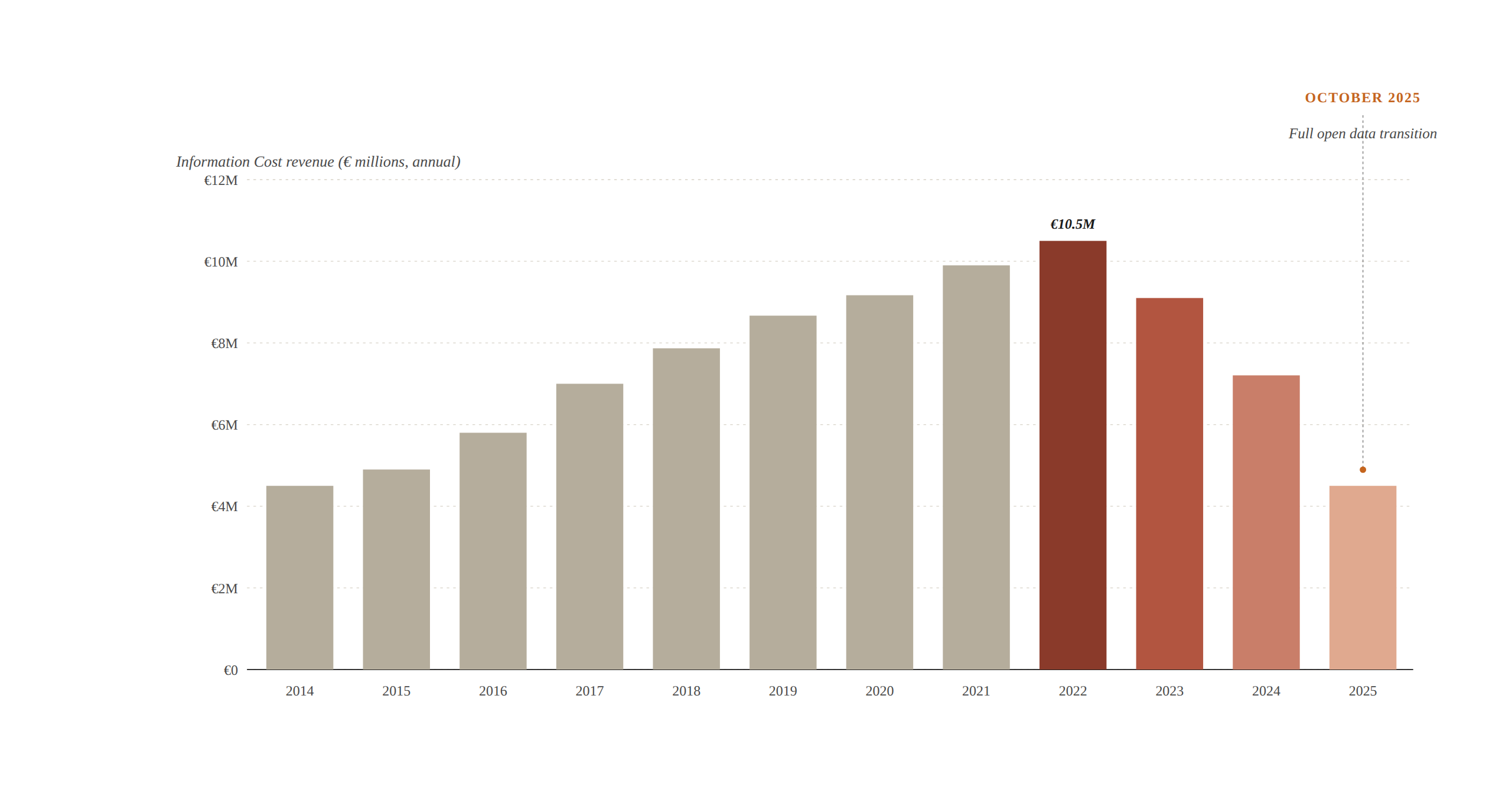}
\caption{Annual Information Cost revenue generated through ECMWF data licensing agreements, 2014--2025 (\euro135 million).}
\label{fig:revenue}
\end{figure}

The client base served under this model was diverse and international. By the time the transition was completed, ECMWF had agreements with organisations spanning energy and utilities, professional weather services, insurance and reinsurance, maritime operations, commodities trading, technology development, and sports.

\begin{figure}[htbp]
\centering
\includegraphics[width=\linewidth]{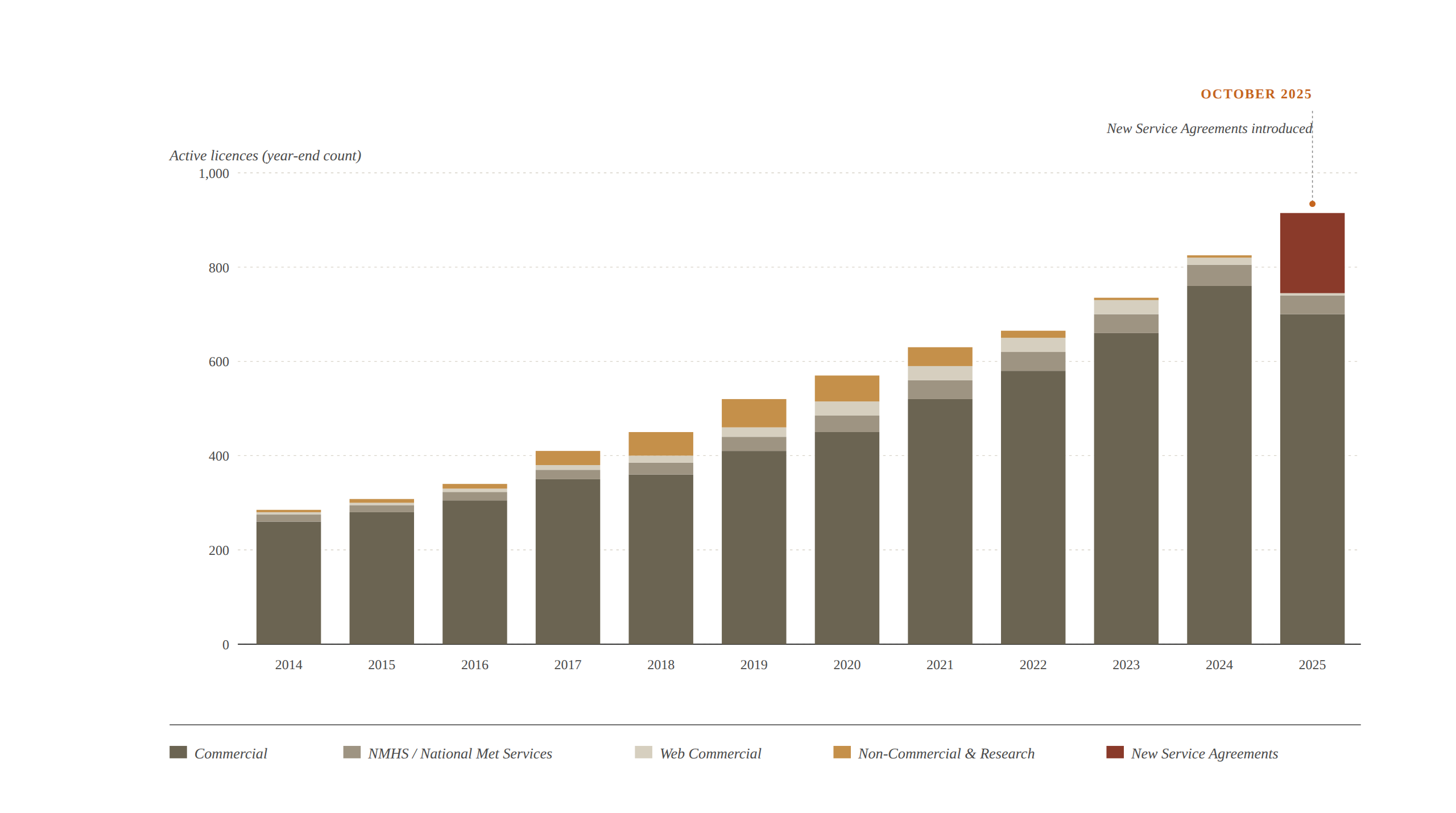}
\caption{Number of active ECMWF data licences by category, 2014--2025.}
\label{fig:licences}
\end{figure}

The policy context that ultimately made the status quo untenable was substantially shaped by European legislative developments. Although ECMWF is not an EU institution, the majority of its Member States are EU members, and its operational activities are deeply embedded in EU frameworks. The EU's Open Data Directive \citep{eu2019} established a clear expectation that public sector information should be made available for reuse, preferably at no cost. The European Strategy for Data \citep{ec2020} reinforced this direction, framing open data as infrastructure for the broader data economy. ECMWF's role as implementing body for Copernicus Climate Change Service (C3S) and Copernicus Atmospheric Monitoring Service (CAMS) further anchored it within an open data framework, as Copernicus data had been freely available since 2015, creating an internal inconsistency in which European climate data was open while operational forecast data from the same institution remained behind a paywall. In December 2019, ECMWF's Council took the formal decision to move gradually to a full open data policy. The regulatory trajectory has since accelerated: the EU's Implementing Regulation on High Value Datasets \citep{ec2023}, which entered into force in June 2024, designates meteorological data as a category that must be made available free of charge by all EU public bodies, extending the open-access obligation well beyond ECMWF to national meteorological services across Europe.

Open-access mandates set the direction of travel, but they do not specify how large operational datasets should be delivered in practice. ECMWF's design choices were shaped by this policy context and by physical constraints: daily production volumes exceeding 300 terabytes, established service obligations to Member and Co-operating States, and finite delivery capacity. The tiered service model emerged from this interplay. It enables compliance with open-access requirements by making core data freely available while retaining cost-recovered operational delivery for users who need 24/7 operational support, customisable service interfaces, and scheduling flexibility. The model is therefore best understood as a means of implementing a controlled open-access policy under realistic infrastructure and support constraints rather than as a departure from it.

%%%%%%%%%%%%%%%%%%%%%%%%%%%%%%%%%%%%%%%%%%%%%%%%%%%%%%%%%%%%%%%%
\section{The Tiered Service Model: Design and Rationale}\label{sec:model}

The central challenge ECMWF faced in designing its open data transition was not political but physical. ECMWF's Integrated Forecasting System (IFS), a state-of-the-art numerical weather prediction model that generates global forecasts between two weeks and seven months ahead, and its Artificial Intelligence Forecasting System (AIFS) currently produce more than 400 terabytes of data per day, of which approximately 100 terabytes are disseminated to Member State users and those with a Service Agreement. This volume grows continuously as new models are added, spatial resolution increases, and the parameter set expands. Providing unrestricted access to the full production volume is not a realistic proposition because the infrastructure costs of delivering data at that scale to an unconstrained global user base would be unsustainable, and for most users, practically unusable.

\begin{figure}[htbp]
\centering
\includegraphics[width=\linewidth]{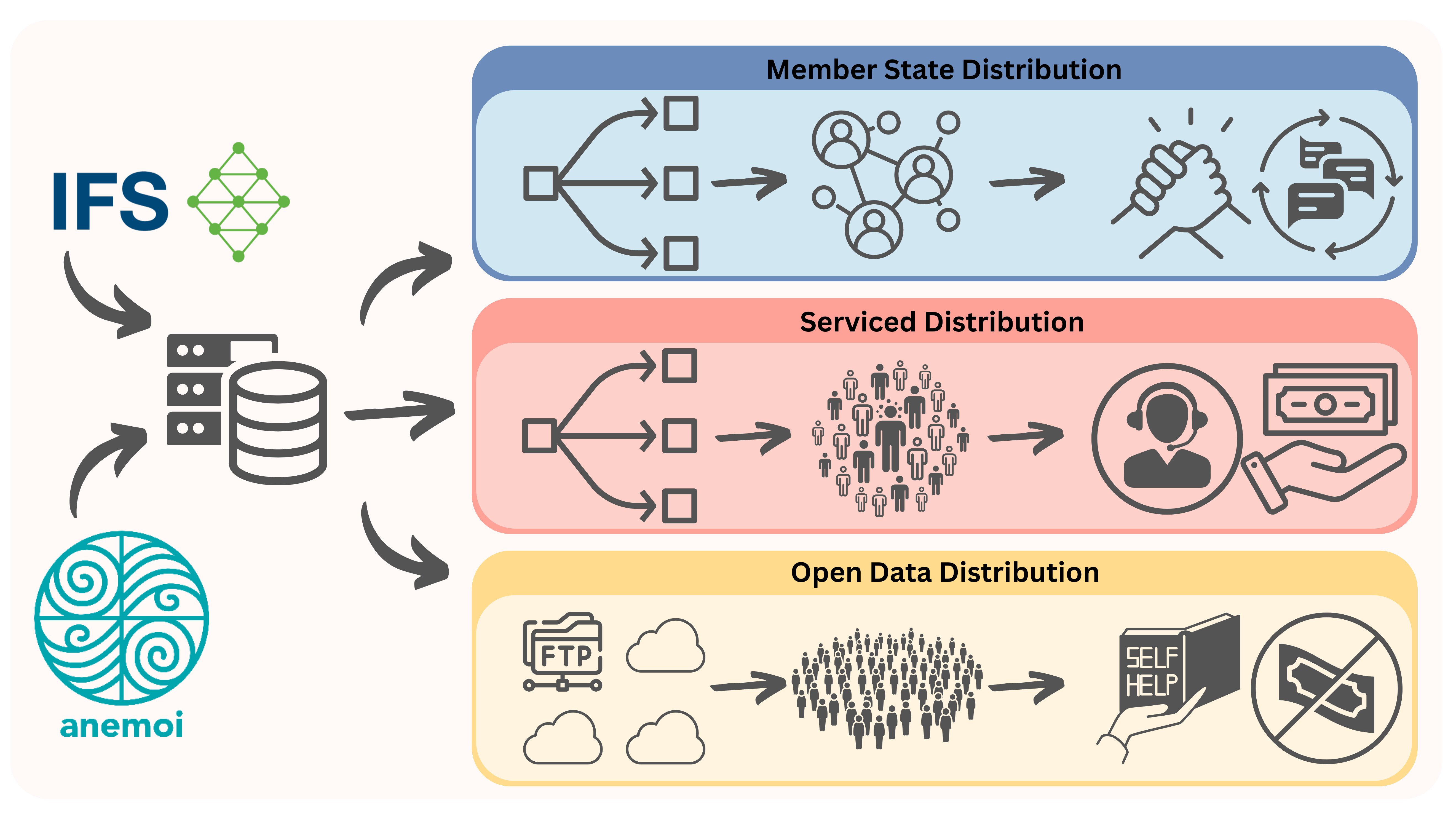}
\caption{Overview of ECMWF's tripartite data distribution model: Member State Distribution (Convention obligation), Serviced Distribution (cost-recovered Paid Service Agreements), and Open Data Distribution (free, CC-BY-4.0).}
\label{fig:tripartite}
\end{figure}

\subsection{Member \& Co-operating State, WMO, and Non-Commercial Access}\label{sec:msaccess}

The foundation of ECMWF's data distribution model is its Convention obligation to Member and Co-operating States. This obligation predates the open data transition and was unaffected by it; a core design principle of the transition was that nothing in the new model should diminish the priority access and dedicated support that Member and Co-operating States have always received.

In compliance with the ECMWF Convention, Member and Co-operating States receive a privileged set of services spanning dedicated data delivery, shared computing infrastructure, archive access, and early access to new products. In practice, this includes the following. Under the Convention, Member and Co-operating States receive one-to-one data delivery tailored to their operational requirements, access to the European Weather Cloud (a shared computing environment for meteorological applications), broader MARS data retrieval access including allocated storage (MARS, the Meteorological Archival and Retrieval System, is ECMWF's operational archive of forecast and observational data), and access to ECMWF's high-performance computing (HPC) environment and file system infrastructure (ECFS). Member and Co-operating States also receive early and privileged access to new and experimental products, including AI-based forecasting models, the Anemoi AI tooling framework (ECMWF's open-source machine learning toolkit for weather forecasting), and pre-operational datasets.

Beyond its own Member and Co-operating States, ECMWF provides a structured set of access provisions for the global meteorological community represented by the World Meteorological Organization (WMO). WMO Members benefit from free access to ECMWF's Open Charts (pre-made meteorological visualisations) and one complimentary ecCharts account per organisation (ecCharts is ECMWF's interactive web-based chart tool). A dedicated WMO Support Dataset has been developed as part of the Systematic Observations Financing Facility (SOFF) initiative, a joint effort by WMO, UNEP, and partner organisations to strengthen the global observing network in developing countries. WMO Members are also entitled to a 50\% discount on Paid Service Agreement costs.

\subsection{The Open Data Tier: Scope and Evolution}\label{sec:opentier}

The free and open data tier was designed to be useful while staying within ECMWF's sustainable capacity. The initial dataset was offered at 0.4-degree spatial resolution (approximately 44 kilometres), broadly equivalent to the resolution of ERA5 reanalysis data and aligned with the WMO Unified Data Policy \citep{wmo2021} categories for the free international exchange of Earth system data. At this resolution, the open data volume was well under 1 terabyte per day.

The dataset was progressively extended to incorporate parameters aligned with the WMO Unified Data Policy framework, with spatial resolution increased to 0.25 degrees (approximately 28 kilometres), sufficient to initialise a Limited Area Model, such as WRF or ICON. Parameters have been added iteratively in response to user feedback and to support the growing use of ECMWF's AIFS. From 2026, spatial resolution will increase further to 0.1 degrees (approximately 9 kilometres), which represents the highest resolution at which the open data tier will be provided and is the native resolution of ECMWF's current operational IFS. A modest reduction in bits-per-value (the precision at which each numerical value is encoded, traded off against file size) will manage the resulting data weight.

Prior to October 2025, the open data tier was provided with a one-hour delay relative to the operational dissemination schedule. Following the full transition, this delay was removed. Service Agreement holders in the Gold and Silver tiers can also arrange delivery ahead of the public dissemination schedule by pre-configuring data requests, enabling data to be pushed to their systems as soon as each forecast cycle completes, rather than at the public release time---a configuration advantage rather than privileged early access to the data itself. In the case of AIFS, which runs on a different computational cycle from IFS, outputs become available earlier in the day than equivalent IFS fields; this earlier availability is offered as a premium feature within the Paid Service Agreement framework.

\begin{figure}[htbp]
\centering
\includegraphics[width=\linewidth]{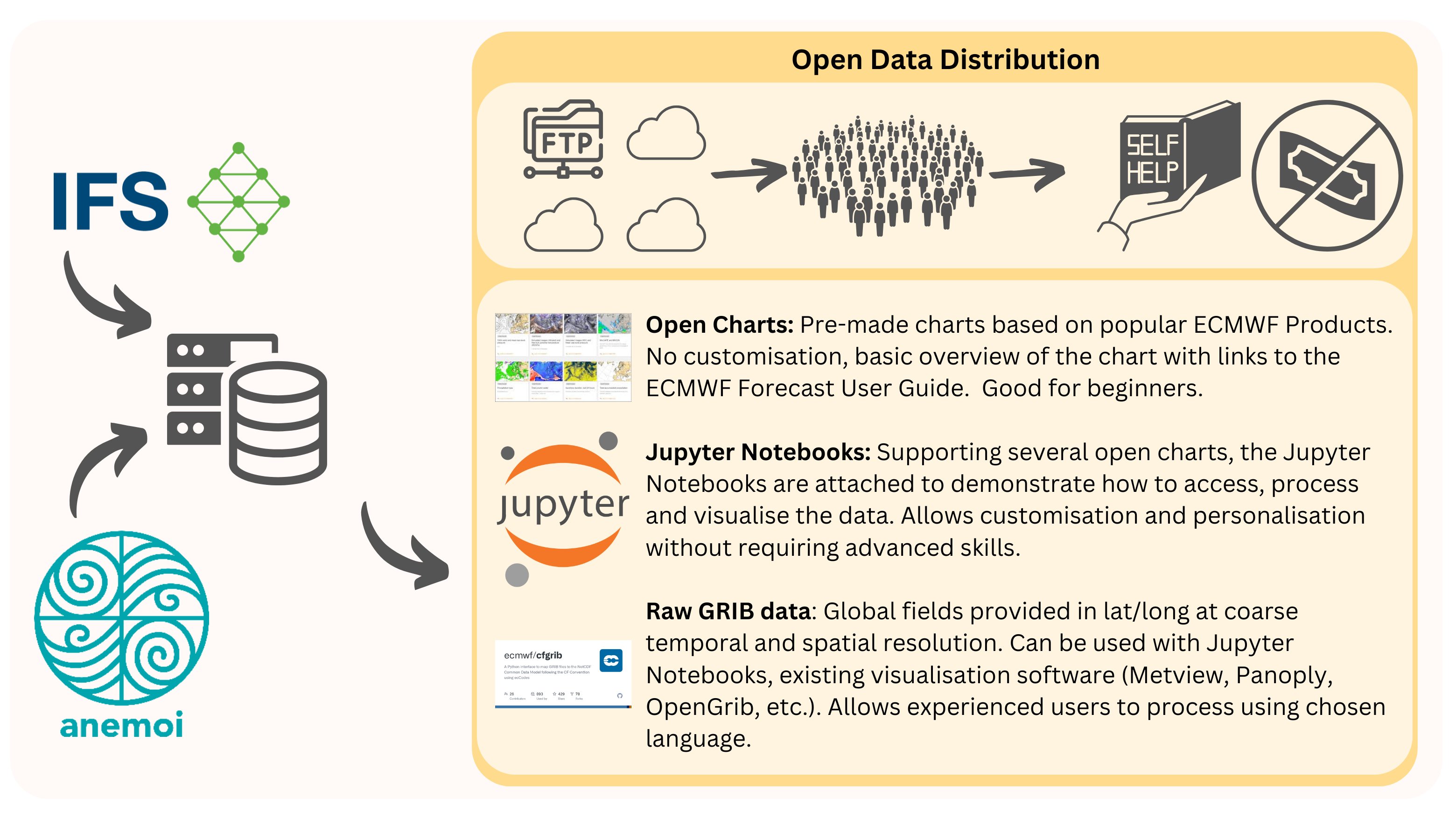}
\caption{Open Data Distribution channel in detail.}
\label{fig:opendata}
\end{figure}

The open tier is complemented by a suite of free supporting resources: pre-made Open Charts for non-technical users who need ready-made visualisations without programming; Jupyter Notebooks for users with moderate technical skills who wish to customise and process the data; and raw GRIB data (a standard meteorological data format) for experienced users processing data programmatically. Free eLearning resources support onboarding across all user types.

\subsection{The Serviced Distribution Tier}\label{sec:servicedtier}

For users whose operational requirements exceed what the open tier provides, ECMWF offers a Serviced Distribution model structured around four bands: Basic, Bronze, Silver, and Gold. This corresponds to the `freemium' archetype identified in public sector information business model taxonomies \citep{ferro2013}: tiers are differentiated not by access to the underlying data, which remains the same IFS output available to all users, but by the level of service, support, and operational flexibility provided.

The Basic tier offers standard delivery on ECMWF's schedule, limited configuration changes per year, and business-hours support. Bronze adds 24/7 support and backup delivery infrastructure. Silver extends this with service interface access and the ability to submit data requests outside standard scheduling. Gold provides full 24/7 autonomy over data configuration, access to MARS via API, ecCharts access, and pre-schedule delivery. Volume is charged separately, with the top band supporting delivery of up to eight terabytes per day.

\begin{figure}[htbp]
\centering
\includegraphics[width=\linewidth]{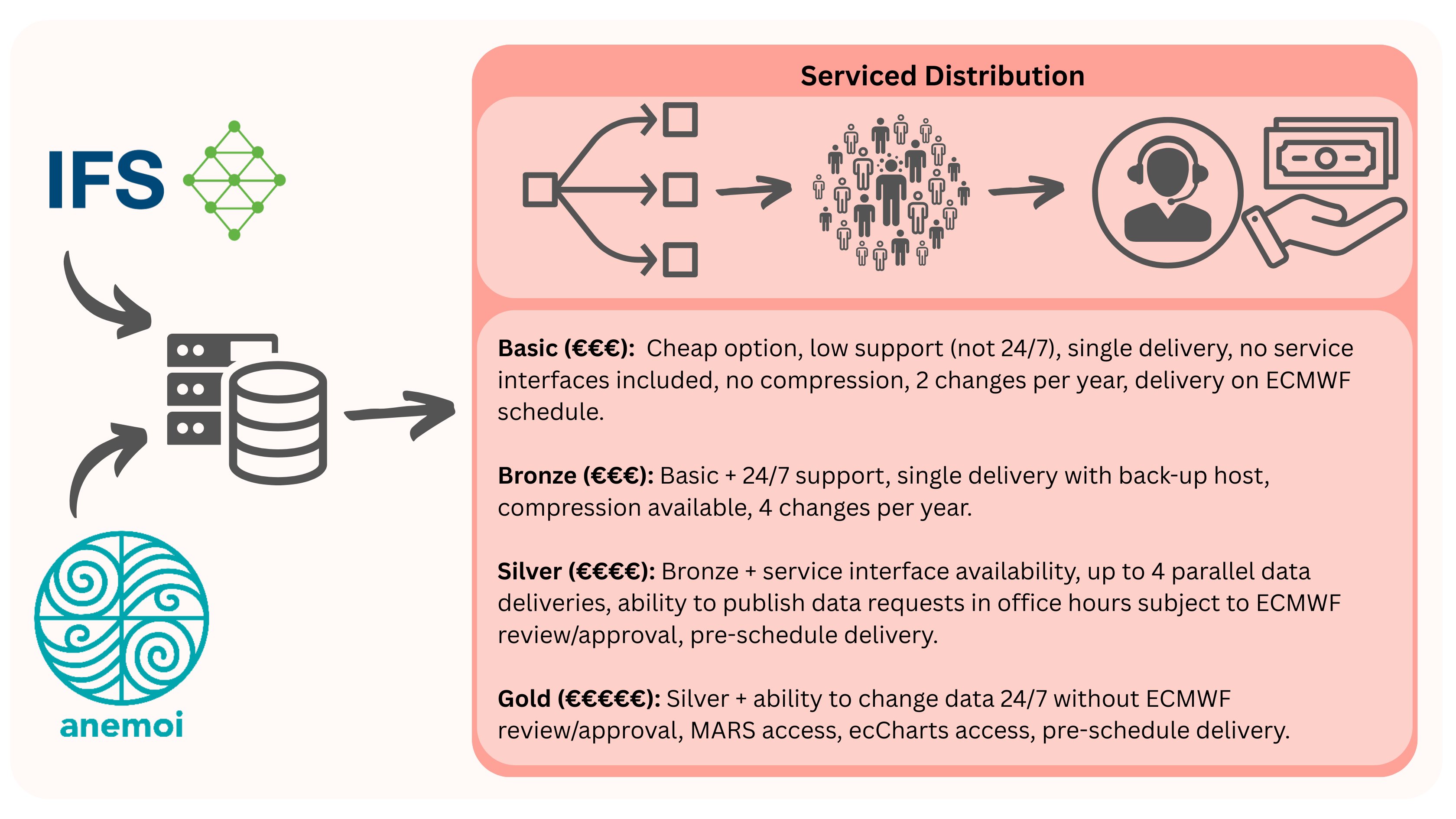}
\caption{Paid Service Agreement channel in detail. All four tiers (Basic, Bronze, Silver, and Gold) carry a service charge; tiers are further differentiated by service level, support availability, delivery flexibility, and volume capacity.}
\label{fig:psa}
\end{figure}

This structure reflects a principle found across a range of publicly funded services: the underlying resource is provided as a public good, while premium service levels are offered on a cost-recovery basis to those whose operational requirements justify the cost \citep{pollock2009,koski2011}. As a comparison, a national health service provides universal access to medical care while private provision offers customised delivery and enhanced service levels; the existence of a premium tier does not diminish the public tier but helps sustain the infrastructure on which both depend \citep{lindsey2006,oecd2009}. This logic holds only where the public tier is genuinely adequately resourced; open data policy frameworks have explicitly cautioned against designs that render the free offer substantively inferior to drive users toward paid alternatives \citep{eu2019}. ECMWF's tiered model operates on the same logic, with Paid Service Agreement revenue contributing directly to the operational budget that sustains forecast quality for all users.

Empirical analysis of UK organisations operating open data business models finds that the Open-Closed Hybrid and Freemium archetypes are associated with statistically significantly higher financial stability than other models, including purely free provision \citep{carbonara2025}. ECMWF's Paid Service Agreement model combines elements of both: data is provided freely under an open licence, while premium delivery services are cost-recovered. The early retention evidence reported in Section~\ref{sec:outcomes} is consistent with this finding.

It is also worth noting that the relationship between tiers is not simply hierarchical. Investments made to serve Member and Co-operating States, in high-performance computing, data retrieval infrastructure, and AI tooling, create capabilities that flow through to all service tiers. Equally, the self-service tools and documentation developed to manage external user demand reduce operational complexity across the board.

%%%%%%%%%%%%%%%%%%%%%%%%%%%%%%%%%%%%%%%%%%%%%%%%%%%%%%%%%%%%%%%%
\section{The Transition Process: An Iterative Approach}\label{sec:transition}

The transition from the licensed model to full open access was not a single policy event but an iterative process spanning five years, from the Council decision in December 2019 to the completion of the transition in October 2025. Each year, ECMWF set a target for revenue reduction, an amount by which Information Cost income would be allowed to fall, and used this as the basis for a structured planning cycle, aligned with ECMWF's annual budget decision process.

It is important to note that the transition was not solely a technical or financial exercise but a political one. The political dimension of the transition was not primarily one of resistance to open data. Member States were broadly supportive of the direction and, in many cases, had advocated for it. The more substantive negotiation concerned funding: each annual step required Member States to accept that reduced Information Cost revenue might necessitate increased national contributions, and few were willing to commit to that in advance of observed outcomes. The iterative structure of the transition reflected this consultative context: it allowed incremental demonstration that the financial impact was within acceptable bounds, supporting informed decision-making by contributing governments about the pace of change. Consensus was therefore cumulative rather than contested: the destination was rarely in doubt; the negotiation concerned the allocation of transition costs and the pace at which uncertainty could be resolved. This dynamic, in which an institution is expected to expand public access while funding the transition without additional burden on contributors, is likely to recur in analogous institutional settings.

The annual cycle proceeded as follows. Financial modelling was used to determine which parameters and resolution improvements could be added to the open tier while remaining within the target revenue reduction. Infrastructure requirements for the expanded open tier were then estimated and provisioned. The updated open tier was implemented, and usage and revenue outcomes were assessed against projections. This assessment fed directly into the following year's planning cycle, with observed user behaviour informing revised assumptions about uptake and retention.

\begin{figure}[p]
\centering
\makebox[\textwidth][c]{%
  \rotatebox{90}{%
    \begin{minipage}{\textheight}
      \centering
      \includegraphics[width=0.9\linewidth]{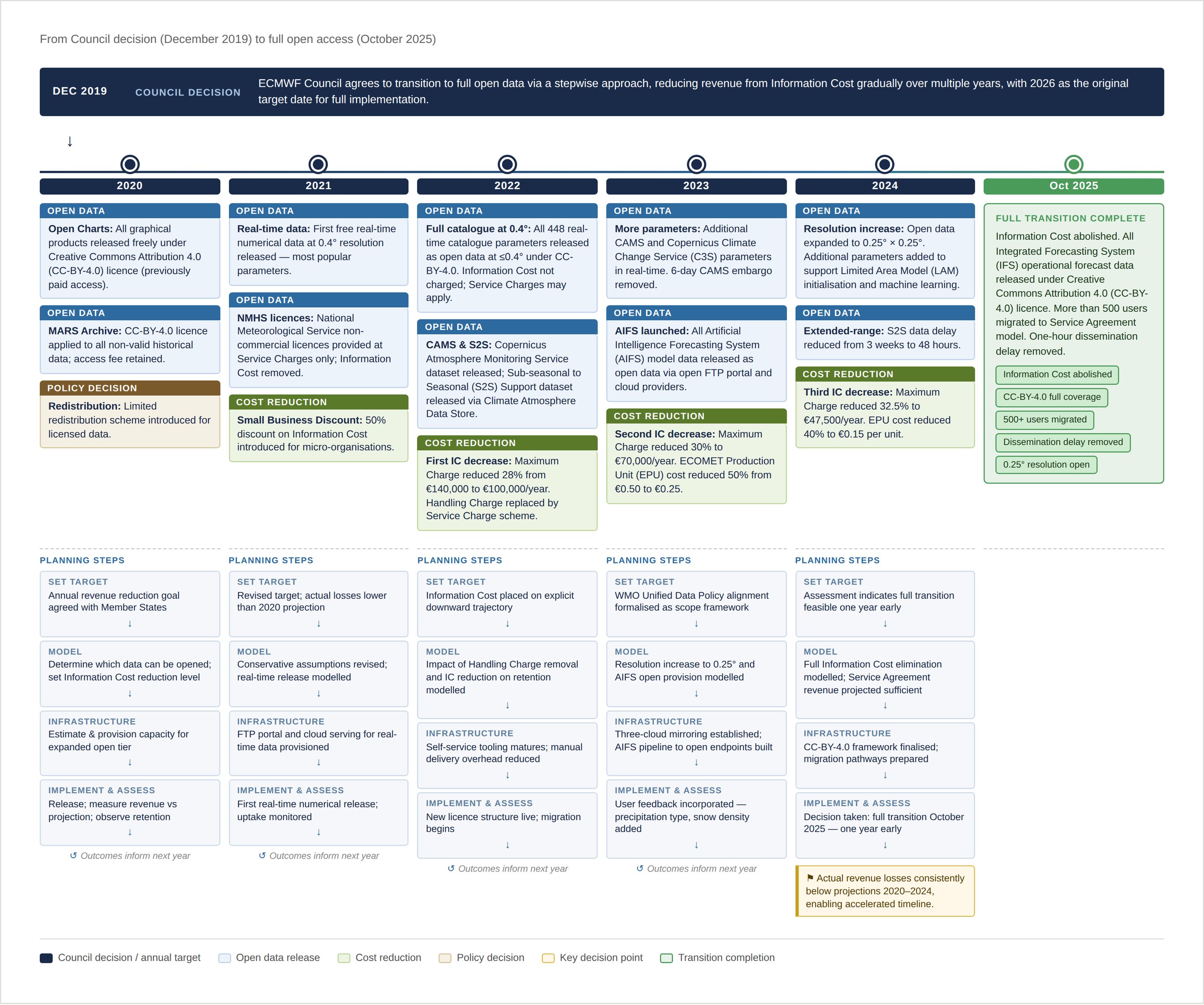}
      \caption{The iterative annual planning cycle, December 2019 to October 2025.}
      \label{fig:cycle}
    \end{minipage}%
  }%
}
\end{figure}

Throughout this process, a consistent finding was that actual revenue losses were smaller than projected. Conservative assumptions about user behaviour, particularly about how many organisations would exit paid agreements entirely once free access became available, proved overly pessimistic. This pattern of over-prediction recurred each year and ultimately led to the decision, taken ahead of schedule, to complete the full transition in October 2025, one year earlier than initially planned.

This iterative structure has several implications for institutions considering similar transitions. It allowed ECMWF to manage financial risk incrementally rather than absorbing the full impact of transition in a single year. It provided a mechanism for learning from observed user behaviour and adjusting assumptions accordingly. And it created a clear, communicable framework for explaining the pace of transition to Member States, whose agreement was required at each stage.

%%%%%%%%%%%%%%%%%%%%%%%%%%%%%%%%%%%%%%%%%%%%%%%%%%%%%%%%%%%%%%%%
\section{Challenges, Trade-offs and Lessons Learned}\label{sec:lessons}

Between the Council decision of December 2019 and the completion of the transition in October 2025, ECMWF navigated a sustained set of operational, financial, and institutional challenges. This section documents those challenges and identifies areas where trade-offs were required.

\subsection{Defining the Open Tier}\label{sec:defining}

Perhaps the most consequential challenge was the question of scope: \emph{what should the free and open data tier include?} Including too little would render the open tier tokenistic; including too much would make the data volume unmanageable for many users, particularly those with limited connectivity, and would place unsustainable demands on delivery infrastructure.

The WMO Unified Data Policy \citep{wmo2021} provided the most solid basis for these decisions, grounding parameter selection in internationally recognised standards. This alignment also served a practical function: it provided an externally referenced framework for explaining the boundaries of the open tier to users who felt their specific requirements had not been met. In practice, the WMO framework provided a floor rather than a ceiling, and decisions about what to include beyond that floor required iterative negotiation between user needs, infrastructure constraints, and the evolving requirements of AI-based forecasting systems.

A related observation is that open data need not mean complete data. The free tier is provided at a lower resolution than the full operational output. The aim is to find the right balance between genuine usefulness and what can be reasonably, reliably, and easily accessed by a wide range of users, including those with limited computational or bandwidth resources. Establishing and communicating this principle early is important to the long-term sustainability of any open data model.

\subsection{Financial Forecasting Under Uncertainty}\label{sec:forecasting}

Each year of the transition period, ECMWF produced revenue forecasts modelling the anticipated impact of Information Cost reductions. These forecasts were shared with Member and Co-operating States to inform their understanding of potential increases in contributions. Each year, those forecasts predicted greater losses than materialised.

This was not a failure of financial modelling but a reflection of genuine uncertainty: there was no comparable precedent, and the behavioural responses of users could only be estimated. In practice, many organisations that had previously paid for data access chose to reinvest savings from eliminated Information Cost into higher service tiers, obtaining more data, faster delivery, or enhanced support rather than simply reducing expenditure. Of the more than 480 organisations that underwent the transition, fewer than 25 discontinued their relationship with ECMWF entirely. This retention rate suggests that institutions contemplating similar transitions may be more financially resilient than worst-case projections imply, though the result reflects specific characteristics of ECMWF's user base---price-sensitive but timeliness-dependent, reliant on real-time access for operational advantage, and operating in a market with few substitute products at equivalent global resolution and latency---and should not be assumed transferable without qualification. Theoretical models of open data business sustainability \citep{carbonara2025} identify cost-recovered service provision as one of the more stable archetypes; ECMWF's experience provides early empirical support for this view.

\subsection{The Redistribution Problem}\label{sec:redistribution}

For much of the licensing era, the prohibition on redistribution was among the most administratively burdensome provisions of ECMWF's data agreements. The boundary between legitimate value addition and redistribution was contested and frequently tested. The transition to CC-BY-4.0 licensing dissolved this problem: attribution is required, but redistribution is permitted, and the compliance overhead has effectively disappeared.

The redistribution activity that has emerged under open access has been largely technical in character. The primary redistributors are communities experimenting with new data formats and access paradigms---Zarr, Analysis-Ready Cloud-Optimised (ARCO) datasets, and API-based services that repackage ECMWF data for specific use cases. These activities are consistent with expected downstream use of open data and, notably, reduce data access friction for a broader range of users. Observed redistribution to date has been predominantly technical (e.g., new formats and access methods), with limited evidence of large-scale reselling.

Several structural factors help explain why large-scale commercial redistribution has not emerged. Matching the reliability and latency of ECMWF's own operational dissemination requires infrastructure investment comparable to that of a national meteorological service, which is rarely justified by a pure redistribution business case. The cloud providers hosting ECMWF mirrors operate under a different economic logic: their incentive is to attract compute and storage customers to their platforms, not to monetise the data itself, which they provide at no cost to end users. And many operational users prefer the authoritative source---both for data provenance and for long-term assurance of continued availability---over third-party redistributors whose longevity is uncertain. Together, these factors mean that CC BY 4.0 open access has not resulted in disintermediation of ECMWF's role.

\subsection{Infrastructure Pressure and Scalability}\label{sec:scalability}

Download volumes from ECMWF's own distribution endpoints increased substantially following the October 2025 transition, from approximately 980 TB/month to 1{,}300 TB/month in the first three months. However, this figure substantially understates the true scale of open data consumption: the majority of downloads occur via cloud mirror platforms, which do not report to ECMWF's own server logs.

ECMWF currently mirrors its open data across three major public cloud platforms. Azure has hosted a mirror since December 2022, serving approximately 130--150 TB/month at peak (2023), and remains active. AWS became the primary mirror platform from early 2025, serving an average of approximately 1{,}230 TB/month across 2025. Following the full open data transition in October 2025, AWS mirror traffic averaged approximately 1{,}530 TB/month, rising to over 1{,}850 TB/month by February 2026. \textbf{Mirror traffic therefore substantially exceeds ECMWF-owned endpoint traffic}: in the post-transition period, AWS alone serves approximately 20\% more data per month than ECMWF's own infrastructure. Google also hosts a mirror, though data from that platform are not included here. Combined with the Copernicus Data Store (CDS), which independently serves approximately 100 TB/day ($\sim$3{,}000 TB/month) of ERA5 and other ECMWF datasets, the total volume of open provision substantially exceeds what ECMWF disseminates to paying Service Agreement holders and Member States ($\sim$100 TB/day). ECMWF is grateful to those cloud providers for their partnership in making this distribution possible.

\begin{figure}[htbp]
\centering
\includegraphics[width=\linewidth]{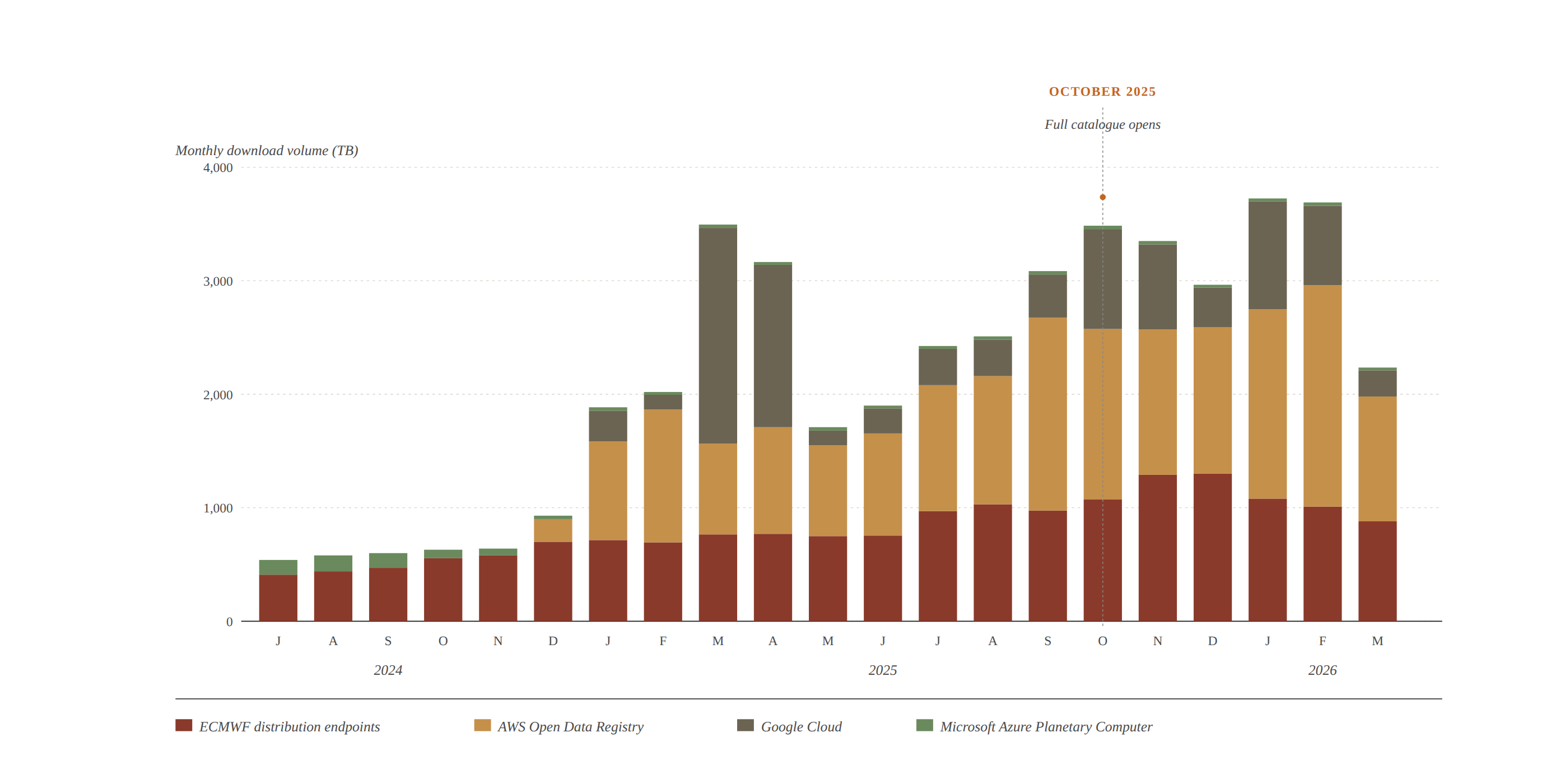}
\caption{Monthly download volume (TB) across ECMWF distribution endpoints and cloud mirrors (AWS Open Data Registry, Google Cloud, Microsoft Azure Planetary Computer), July 2024 -- March 2026. Bars to the right of the dashed line represent the post-transition period from October 2025.}
\label{fig:downloads}
\end{figure}

This mirroring arrangement is a pragmatic solution to a structural challenge that remains unresolved. As the open data tier expands in resolution and model coverage, the volume of data to be mirrored will grow accordingly. The Copernicus programme's approach to data provision establishes that a copy of data should be freely and openly accessible, while alternative access methods offering enhanced service levels may be subject to charges \citep{ec2021}. ECMWF's approach follows this logic. The physical reality of producing and disseminating more than 100 terabytes of data per day means that some form of access differentiation is not a commercial choice but an infrastructural necessity.

Geographic distribution of usage cannot be fully determined from ECMWF's own server logs, as the majority of downloads occur via cloud mirror endpoints which do not report user location. This is itself a structural limitation of cloud-mirrored open data distribution that merits attention in future infrastructure planning.

\subsection{AIFS and Emerging Sustainability Questions}\label{sec:aifs}

The emergence of ECMWF's Artificial Intelligence Forecasting System (AIFS) as an operational product has introduced a new dimension to the sustainability of the open data model. AIFS outputs are provided freely and openly as soon as they are generated---earlier in the day than equivalent IFS fields, owing to differences in the computational cycle---consistent with ECMWF's open data commitments. However, as AIFS model quality improves, this creates a potential tension with the Paid Service Agreement model: if high-quality AI forecast data is available freely and without delay, the rationale for service agreements covering conventional IFS outputs may be affected. This tension has not yet materialised into a significant problem, but it is foreseeable and will require active management. It is not unique to ECMWF---any operational forecasting centre developing AI capabilities will face the same question. A related consideration is that AIFS itself is open source. Users with access to a modest set of initial conditions---smaller still for AI-based limited area models---can run inference independently, which means the question of how ECMWF provides AIFS output is partly decoupled from whether those outputs are accessible at all. This further shifts the locus of the sustainability question from data access to infrastructure and compute provision.

\subsection{Managing Expectations}\label{sec:expectations}

Open data generates demand that consistently outpaces institutional capacity to provide. Users with access to data at 0.25-degree resolution request 0.1-degree resolution; users with six-hourly data request hourly resolution; users with access to 100 parameters request 200. This is not a criticism of users; it reflects the genuine and growing appetite for high-quality meteorological data across an expanding range of applications---but it creates a sustained management challenge that institutions embarking on open data transitions should plan for explicitly. The emergence of AI-based forecasting, which can produce high-resolution fields from comparatively modest input data, may alter this dynamic---users may increasingly generate locally the resolution they previously requested centrally.

The geographic distribution of demand also raises questions about long-term funding equity. Usage of ECMWF's open data extends well beyond Europe, including regions and economies that make no direct contribution to ECMWF's operational funding. This is consistent with the intended purpose of open data, but it raises questions about the sustainability of a model in which a subset of contributing states bears the full cost of infrastructure used globally. It is also worth recognising the reverse relationship: ECMWF's forecasts depend on a vast global infrastructure of in situ observations, satellites, ship and buoy reports, and aircraft measurements, much of which is contributed freely by national services, operators, and international programmes worldwide. Global meteorology operates on a longstanding principle of reciprocal open exchange, and any discussion of funding equity must acknowledge that ECMWF is both a beneficiary and a provider within that system. These questions do not have easy answers, and they are noted here because any honest account of the open data transition at this scale must acknowledge the funding asymmetry that underpins it.

\subsection{Pressure on Service Charges}\label{sec:pressure}

A small number of users have repeatedly sought reductions in Paid Service Agreement charges, on the basis that the elimination of Information Cost should be extended to all aspects of ECMWF's service provision. This reflects a misreading of the open data model that is worth addressing explicitly.

Service charges are designed to reflect the marginal costs of providing enhanced, reliable, and operationally guaranteed data delivery, and to fund continued investment in the delivery infrastructure and new service capabilities on which future users will depend. Reducing these charges in response to pressure would transfer costs from users to the Member and Co-operating States who ultimately underwrite ECMWF's budget, and who must agree to any change in the financial model. The sustainability of the open data tier depends on the Paid Service Agreement tier remaining viable. Reducing service charges under commercial pressure would over time degrade service quality for all users, ultimately acting against the interests of the paying customers themselves.

\subsection{Limitations of This Account}\label{sec:limitations}

Several limitations of the evidence presented here should be acknowledged explicitly.

The post-transition observation window covers six months (October 2025 -- March 2026). While this is a longer window than many early evidence reports, it remains insufficient to draw firm conclusions about long-term financial sustainability. Early retention rates may reflect contractual inertia or unexpired agreements rather than settled user preference; the more meaningful test will be whether retention holds through two or three full renewal cycles. The authors commit to a follow-up account when 12-month data are available.

Retention is reported as headcount rather than revenue. The paper demonstrates that more than 93\% of previously paying organisations maintained a Paid Service Agreement, but it does not segment by tier, user type, or geography, nor does it report whether organisations migrated to lower tiers. Revenue retention in EUR would be a more meaningful indicator of financial sustainability; this data will be reported in future work.

The cost side of the transition is not reported. ECMWF does not operate cost-based accounting at the level of granularity required to attribute infrastructure, staff, and delivery costs to individual data products or user tiers. This is a structural limitation of the available data rather than an intentional omission, and it means the paper cannot demonstrate net financial sustainability---only that revenue was retained at levels exceeding projections.

The counterfactual is unmodelled. Information Cost revenue was already on a mandated downward trajectory from 2019; some of the observed retention would likely have occurred under any transition pathway. The paper does not attempt to separate the effect of the open data transition from the pre-existing revenue trend.

These limitations do not invalidate the findings but do bound what can be claimed. The paper is best understood as an early evidence report from a transition in progress, not a demonstration of long-term sustainability.

%%%%%%%%%%%%%%%%%%%%%%%%%%%%%%%%%%%%%%%%%%%%%%%%%%%%%%%%%%%%%%%%
\section{Outcomes}\label{sec:outcomes}

The observations reported here draw on internal administrative records covering 1 October 2025 to 31 March 2026, including the agreements ledger, download volumes from ECMWF-owned endpoints, and Data Support tickets. These early results should be interpreted with caution. Several indicators showed variability during the transition period, and in many cases the underlying drivers---such as shifts in user behaviour, organisational risk appetite, or evolving AI-based alternatives---cannot yet be disentangled from short-term adjustment effects. ``Retention'' denotes maintenance of a Paid Service Agreement through the first renewal cycle; ``download volume'' denotes data retrieved from ECMWF-owned endpoints and excludes cloud-mirror traffic.

\begin{itemize}[leftmargin=*]
\item \textbf{User base and download volume.} Open access removed both financial and policy-related barriers to entry for research institutions, national meteorological services, commercial enterprises, academic groups, and individual developers. Beyond licensing fees, the previous framework carried ambiguities about permissible use, particularly around derivation and redistribution, that an open licence resolves unambiguously, unlocking the full potential of the data for downstream applications. Monthly download volumes from the ECMWF open-data endpoint increased from 980 TB/month prior to the transition to a maximum of 1301 TB/month in December 2025. Geographic distribution cannot be inferred from ECMWF-owned logs, as mirrors do not report location.
\item \textbf{Revenue retention.} Paid Service Agreement uptake remained higher than conservative projections throughout the observation period, and continued to grow into Q1 2026. The retention rate exceeded 93\%, with fewer than 25 of more than 480 migrated organisations discontinuing their relationship. Savings from the elimination of the Information Charge were often reinvested in higher service tiers, and new Paid Service Agreement sign-ups accelerated markedly in the first quarter of 2026.
\item \textbf{Operational continuity.} Member and Co-operating States experienced no reduction in service quality or priority access. For Service Agreement holders, migration was managed through defined transition pathways. The volume of transition-related support requests increased initially and then normalised.
\item \textbf{Limitations.} A six-month window is insufficient to infer long-term financial trajectories. Mirror-based distribution limits attribution and demand forecasting. The interaction between freely available AI-based forecasts and paid service tiers warrants continued monitoring.
\end{itemize}

%%%%%%%%%%%%%%%%%%%%%%%%%%%%%%%%%%%%%%%%%%%%%%%%%%%%%%%%%%%%%%%%
\section{Conclusion}\label{sec:conclusion}

The transition documented in this paper was motivated by a convergence of policy pressures, scientific principle, and institutional pragmatism. The EU's Open Data Directive, the European Strategy for Data, and the internal inconsistency of maintaining a restricted licensing model alongside freely available Copernicus products made the direction of travel clear. What was less clear, in December 2019, was whether the transition could be accomplished without compromising the financial sustainability of the operational infrastructure on which the value of the data depends.

The evidence from the first six months following the October 2025 transition is consistent with the hypothesis that a tiered service model can be designed to balance open access with operational funding requirements. Early indicators show increased uptake across a broader user base, encompassing research institutions, national meteorological services, commercial enterprises, and individual developers previously excluded by licensing barriers; revenue retained at levels exceeding conservative projections; and operational continuity for Member and Co-operating States maintained without disruption. Cloud mirror data indicate that total open provision already exceeds paid dissemination volumes, providing concrete evidence of the scale of the open data impact. These outcomes are preliminary, and the limitations of a six-month evidence base are acknowledged. It remains possible that some of the apparent stability observed in the first months reflects temporary conditions, such as heightened user attention, novelty effects, or transitional support structures, that may not persist over longer renewal cycles.

The model itself is not novel. The principle that a publicly funded resource should be universally accessible, while the operational services that deliver it may be subject to cost recovery, is established across a range of public service domains. This paper contributes a detailed account of how that principle was operationalised in the specific context of a large-scale operational meteorological data provider---the design choices made, the iterative process followed, the challenges encountered, the trade-offs accepted, and the outcomes observed.

Several questions remain open. The long-term trajectory of revenue under open access will only become clear over multiple renewal cycles. The attribution and visibility of open data usage remain structurally limited. The emergence of AI-based forecast products as freely available operational outputs raises questions about the long-term rationale for service agreements that will require active management. And the question of funding equity, in which the costs of producing and distributing globally used data are borne by a subset of contributing states, deserves wider attention in international policy discussions about the governance of scientific data infrastructure.

This paper has not attempted to resolve these questions. Its aim has been to document, as honestly as possible, what one institution did, what proved difficult, and what the early evidence suggests about the viability of the approach. The authors hope that this account will be of practical use to institutions that will face similar decisions. For European National Meteorological Services now implementing shared infrastructure to meet the EU's High Value Datasets requirements \citep{ec2023,eumetnet2024}, the evidence presented here may be of particular relevance. We hope it contributes to the evidence base on which a sound open data policy can be built.

\subsection*{Lessons Learned}

\begin{itemize}[leftmargin=*]
\item Tiered service provision offers a financially viable path through the tension between open access mandates and cost-recovery requirements.
\item Revenue loss projections consistently overestimated actual losses; user behaviour proved more resilient than conservative models assumed.
\item Alignment with internationally recognised data policy frameworks (WMO Unified Data Policy) provides a principled and communicable basis for defining the scope of the free tier.
\item Open data need not mean complete or perfect data. The aim is to find the right balance between genuine usefulness and what can reliably be delivered to a wide range of users, including those with limited resources.
\item Self-service infrastructure and documentation are not optional. The volume of demand generated by open access cannot be managed through manual support processes.
\item Downstream technical activity, new tools, formats, and services built on open data, tend to be broader and more varied than anticipated under a licensing model and reduce data access friction for new categories of users.
\item Pressure to extend open data logic to service charges should be resisted on principled grounds: customisable service provision is sustainable only when service charges reflect their marginal cost, and reducing them under commercial pressure would compromise service quality to the detriment of paying customers.
\item The funding equity question---who bears the cost of infrastructure that benefits a global user base---is unresolved and deserves wider policy attention.
\end{itemize}

%%%%%%%%%%%%%%%%%%%%%%%%%%%%%%%%%%%%%%%%%%%%%%%%%%%%%%%%%%%%%%%%
\section*{Acknowledgments}
The authors thank ECMWF's Member and Co-operating States for their drive to open ECMWF data. The authors also thank Amazon AWS, Microsoft Azure, and Google for their partnership in mirroring ECMWF open data, which has been essential to the scale of distribution achieved. Further information about ECMWF's open data offer is available at \url{https://www.ecmwf.int/en/forecasts/datasets/open-data}. The authors also acknowledge the use of Grammarly (\href{https://www.grammarly.com}{grammarly.com}) for consistency of English language across the manuscript, and Claude (Anthropic; \href{https://claude.ai}{claude.ai}) for assistance with reviewing manuscript structure and clarity, and for editor-style review of drafts during preparation. The authors are entirely responsible for the scientific content of the paper and have reviewed and verified all output.

\section*{Funding statement}
This work was carried out as part of ECMWF's operational activities and did not receive specific external funding.

\section*{Competing interests}
All authors are current or former employees of ECMWF, the institution described in this paper. Florence Rabier was Director General of ECMWF during the period in which the open data policy was adopted and is now Non-Executive Director of the UK Met Office. The authors have endeavoured to present the institutional experience as accurately and objectively as possible.

\section*{Data availability statement}
Aggregated data underlying Figures~\ref{fig:revenue}--\ref{fig:licences} and the `Outcomes' section will be provided upon request. These data include: (i) annual information-cost revenue totals and licence counts (2014--2025); (ii) monthly download volumes from ECMWF-owned endpoints (Oct 2025--March 2026); (iii) monthly cloud mirror download volumes from AWS (December 2024--March 2026) and Azure (December 2022--March 2026); and (iv) monthly counts of data-support tickets over the same period. Contract-level records and personally identifiable information cannot be shared due to confidentiality constraints; the deposited aggregates reproduce the results reported in the article.

\section*{Ethical Standards}
This study involves no human subjects. All data used are institutional administrative records held by ECMWF.

\section*{Author contributions}
Conceptualisation: E.P., V.B., U.M., F.P., F.V., F.R. Methodology: E.P., V.B., U.M., F.P., F.V., F.R. Data curation: E.P. Data visualisation: E.P. Writing -- original draft: E.P. Writing -- review \& editing: E.P., V.B., U.M., F.P., F.V., F.R. Supervision: V.B., U.M., F.P., F.V., F.R. All authors approved the final submitted draft.

%%%%%%%%%%%%%%%%%%%%%%%%%%%%%%%%%%%%%%%%%%%%%%%%%%%%%%%%%%%%%%%%
\bibliographystyle{abbrvnat}
\bibliography{refs}

\end{document}